\documentclass[USenglish,twocolumn]{article}

\usepackage[utf8]{inputenc}
\usepackage[big]{dgruyter}

\begin{document}

  \articletype{Research Article{\hfill}Open Access}

  \author[1]{Joris Vos}

  \author[2]{Maja Vu{\v c}kovi{\'c}}

  \affil[1]{Instituto de F\'{\i}sica y Astronom\'{\i}a, Universidad de Valparaiso, Gran Breta\~{n}a 1111, Playa Ancha, Valpara\'{\i}so 2360102, Chile, Chile, E-mail: joris.vos@uv.cl}
  \affil[2]{Instituto de F\'{\i}sica y Astronom\'{\i}a, Universidad de Valparaiso, Gran Breta\~{n}a 1111, Playa Ancha, Valpara\'{\i}so 2360102, Chile}

  \title{\huge Constraining Roche-Lobe Overflow Models Using the Hot-Subdwarf Wide Binary Population}

  \runningtitle{Constraining Roche-Lobe Overflow Models Using the Hot-Subdwarf Wide Binary Population}


\begin{abstract}
{One of the important issues regarding the final evolution of stars is the impact of binarity. A rich zoo of peculiar, evolved objects are born from the interaction between the loosely bound envelope of a giant, and the gravitational pull of a companion. However, binary interactions are not understood from first principles, and the theoretical models are subject to many assumptions. It is currently agreed upon that hot subdwarf stars can only be formed through binary interaction, either through common envelope ejection or stable Roche-lobe overflow (RLOF) near the tip of the red giant branch (RGB). These systems are therefore an ideal testing ground for binary interaction models. With our long term study of wide hot subdwarf (sdB) binaries we aim to improve our current understanding of stable RLOF on the RGB by comparing the results of binary population synthesis studies with the observed population. In this article we describe the current model and possible improvements, and which observables can be used to test different parts of the interaction model.
}
\end{abstract}


  \journalname{Open Astronomy}
\DOI{DOI}
  \startpage{1}
  \received{..}
  \revised{..}
  \accepted{..}

  \journalyear{2017}
 \journalissue{Special issue of the 8th Meeting on Hot Subdwarfs and Related Objects}

\maketitle
\section{Introduction}
The impact of binarity on stellar evolution is an important issue in the final evolution of stars. Depending on the spectral type between 30 and 80 \% of all stars resides in a binary or higher order multiple system \citep{Raghavan2010}. If the initial orbital period ranges between roughly $\sim$ 10 and 1000 days, these systems will interact on the red giant branch (RGB). However, these interaction mechanisms are currently not well understood, and are treated in a parameterised way, see for example \citet{Ivanova2013} for common envelope (CE) ejection and \citet{Vos2015} for stable Roche-lobe overflow (RLOF). 

Hot subdwarf B-type (sdB) stars are formed solely through binary interaction, making them ideally suited to study these binary interaction mechanisms. While short period sdB binaries are the product of CE ejection, wide sdB binaries are formed through stable RLOF near the tip of the RGB \citep{Han2002, Han2003}. Currently 11 wide sdB binaries have solved orbits \citep{Vos2017a}, most of which are the result of a long term observing campaign using HERMES at the 1.2m Mercator and UVES at the 8m VLT telescope \citep{Vos2017b}. Although this is still a small sample, several interesting result have appeared.

In this article we will give a brief overview of the model used to describe stable RLOF, and how the observed population of wide sdB binaries can be used to improve it. To show the importance of studing these interaction mechanisms, in Sect.\,\ref{sec:interaction} the fraction of low mass binaries that will interact on the RGB is estimated. In Sect.\,\ref{sec:model} the model for RLOF in MESA is described, and in Sect.\,\ref{sec:improvements} improvements for this model, and observables related to the different processes are discussed.

\section{Interaction on the RGB}\label{sec:interaction}

\begin{figure*}
\includegraphics{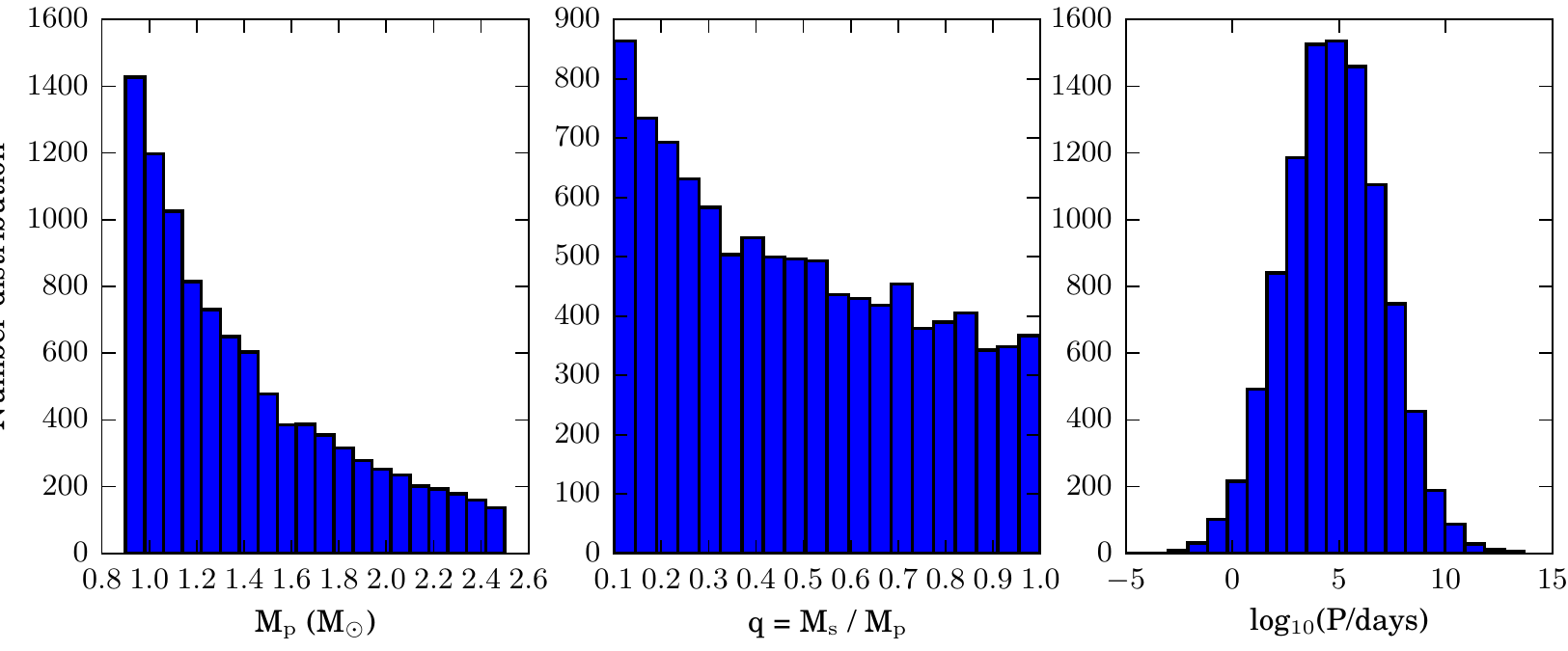}
\caption{The primary mass (left), mass-ratio (middle) and orbital period distribution (right) used when estimating the fraction of low mass binaries that will undergo mass-loss on the RGB. The primary- mass distribution is the Salpeter IMF, the mass-ratio distribution is based on observations of G-A type binaries in the solar neighbourhood, and the orbital-period distribution is based on observations of G-type binaries in the solar neighbourhood.}\label{fig:distributions}
\end{figure*}

To estimate how many low mass binary systems will interact on the first giant branch we have to take into account the distribution of the masses, mass ratios and orbital periods of the low-mass-binary population. In the following we focus solely on binaries containing a primary with a mass between 0.9 and 2.5 M$_{\odot}$. To estimate the mass distribution we use the Salpeter initial mass function (IMF, \citealt{Salpeter1955}), the mass-ratio distribution follows an exponential decaying function, and is roughly independent of mass as estimated from observations of G to A-type binaries in the solar neighbourhood by \citet{Metchev2009, deRosa2012} and \citet{Goodwin2013}. In this case we have limited the mass-ratios between $q = M_{\rm s}/M_{\rm p}$ = 0.1 - 1.0. The orbital period distribution depends strongly on the stellar type, and varies from a log normal distribution centered around $\log_{10}{P/{\rm days}} \approx 4.8$ for G-type star in the solar neighbourhood \citep{Duquennoy1991} to an exponentially decaying distribution for binary with heavy (M$\sim$16 M$_{\odot}$) primaries \citep{Sana2012}. As our focus is on the low mass stars we have used the log normal distribution of \citep{Duquennoy1991}. Our sample is then defined by the following distributions:
\begin{align*}
 M_{\rm p} &\sim M^{-2.35}, \\
 q &\sim q^{-0.4}, \\
 \log{P} &\sim \exp \left( \frac{-(\log{P} - 4.8)^2}{2 \cdot 2.3^2} \right),
\end{align*}
which are shown in Fig.\,\ref{fig:distributions}. 

To check if these systems will eventually interact on the RGB, the Roche-lobe radius of the primary is calculated and compared with the expected radius at the start and at the end of the RGB. To estimate the radius of the RGB, MESA \citep{Paxton2015} models obtained from the MESA Isochrones \& Stellar Tracks database\footnote{\url{http://waps.cfa.harvard.edu/MIST/index.html}} (MIST, \citealt{Dotter2016}) were used. For the different masses, a stellar evolution track calculated for solar metallicity was used to estimate the radius at the end of the main sequence and at the end of the RGB. If the stellar radius would exceed the Roche-lobe radius, it will start mass-loss. We find that 2\% of the systems will interact on the MS, while another 15\% will start mass-loss on the RGB. While this is a very rough calculation, it is based on properties of the observed binaries in the solar neighbourhood, and in that sense provides a good estimate of the percentage of binaries that will undergo a binary interaction phase while ascending the RGB.

\section{RLOF model}\label{sec:model}

\begin{figure*}
\includegraphics{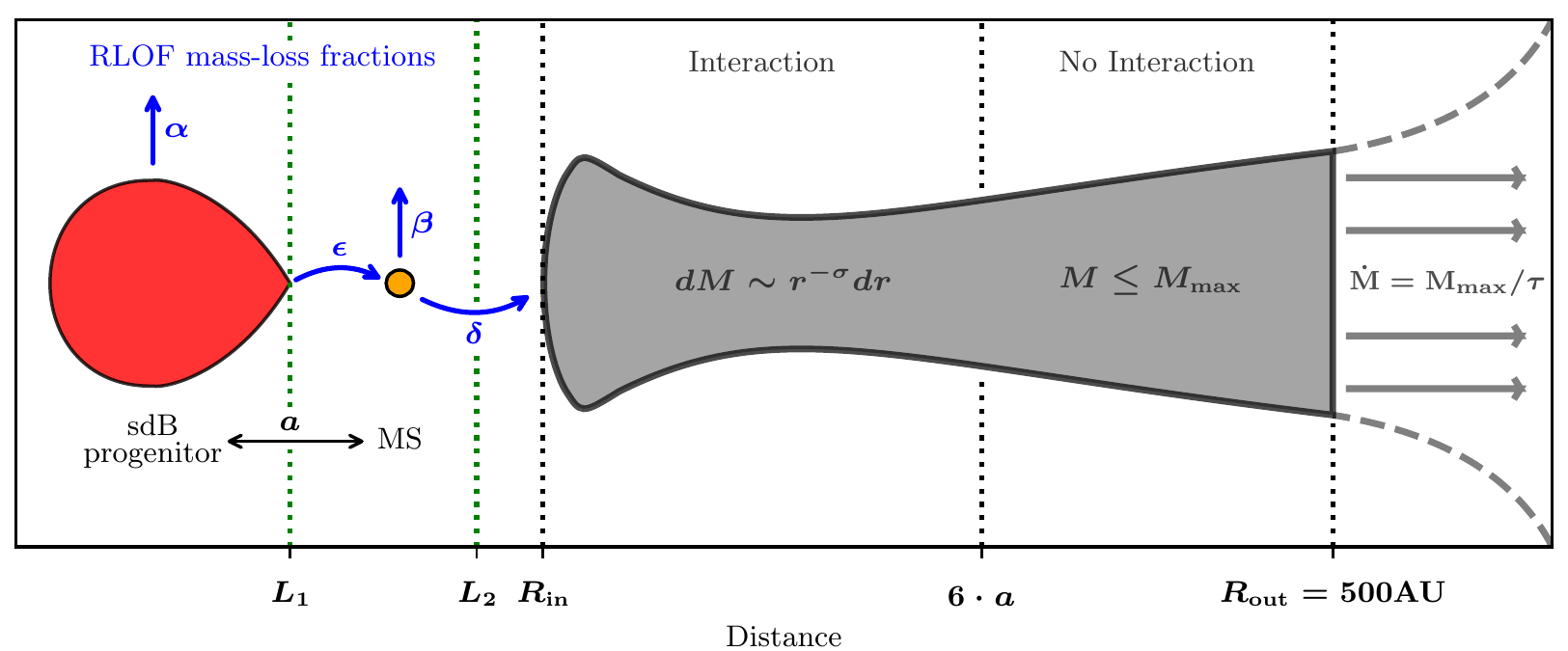}
\caption{The Roche-lobe overflow model used in MESA, including the formation of a circumbinary disk.}\label{fig:RLOF_in_MESA}
\end{figure*}

The model that is used to describe RLOF, including the formation of a circumbinary (CB) disk in MESA is described in detail in the Appendix of \citet{Vos2015}. A visual representation of this model is shown in Fig.\,\ref{fig:RLOF_in_MESA}. The mass-loss rate is calculated in the standard way, based on the Roche-lobe overflow fraction using the prescription of \citet{Ritter1988} and \citet{Kolb1990}. What happens with that lost mass is parameterised using the mass-loss fraction of \citet{Tauris2006}. This system describes four mass-loss fractions ($\alpha, \beta, \delta, \epsilon$), which are shown in blue on Fig.\,\ref{fig:RLOF_in_MESA}:
\begin{itemize}
 \item $\alpha$: mass lost from the vicinity of the donor as a fast wind (Jeans mode). This is modelled as a spherically symmetric outflow from the donor star in the form of a fast wind. The mass lost in this way carries the specific angular momentum of the donor star.
 \item $\beta$: mass lost from the vicinity of the accretor as a fast wind (Isotropic re-emission). A flow in which matter is transported from the donor to the vicinity of the accretor, where it is ejected as a fast isotropic wind. Mass lost in this way carries the specific angular momentum of the accretor.
 \item $\delta$: mass lost through the outer Lagrange point (L2) which is modelled as a circumbinary coplanar toroid. This toroid is placed at a distance $R_{\rm toroid} = 1.2^2 \cdot a$, where $a$ is the semi-major axis, to correspond with the radius of L2.
 \item $\epsilon$: mass transferred through the inner Lagrange point (L1) and accreted by the companion ($\epsilon = 1 - \alpha - \beta - \delta$).
\end{itemize}

To explain the eccentricity of the observed wide sdB population, two eccentricity pumping mechanisms are incorporated in the model: phase-dependent mass-loss during RLOF, where the difference between the mass-loss rate near apastron versus the rate near periastron can increase the eccentricity of the orbit \citep{soker2000, Eggleton2006}. The second mechanism is eccentricity pumping though Lindblad resonances between a CB disk and the binary system. This disk is formed by the mass lost from the donor during RLOF though L2 (mass-loss fraction $\delta$). The interaction between the disk and the binary is modelled using the description of \citet{Artymowicz1994} and \citet{Lubow1996}. The CB disk and its evolution is implemented in a parameterised way using the parameters indicated in grey on Fig.\,\ref{fig:RLOF_in_MESA}:
\begin{itemize}
 \item R$_{\rm in}$: the inner radius of the CB disk. It is implemented using the fitting formula derived by \citet{Dermine2013} based on the results of smooth particle hydrodynamical (SPH) simulations by \citet{Artymowicz1994}.
 \item R$_{\rm out}$: the outer radius of the CB disk.
 \item $\sigma$: the mass distribution constant. The surface mass distribution of the disk is described as $dM/dA \sim r^{-\sigma} dr$.
 \item M$_{\rm max}$: the maximum mass in the disk at any given time. If more mass would be accreted in the disk, it is lost immediatly from the system. 
 \item $\tau$: the maximum lifetime of the disk. $\tau$ defines how long the disk will last when no more mass is accreted into it from the binary system. The mass-loss rate from the disk to infinity is then defined as $\dot{M} = M_{\rm max} / \tau$.
\end{itemize}

The surface mass distribution of a CB disk is usually modelled with $\sigma = 1$, based on observations of the inner parts of proto planetary disks. The mass distribution of a CB disk is not constant, and $\sigma$ actually increases with radius, but as only the inner part of the disk is important for the interaction with the binary, this radius dependence is ignored. Furthermore, the total mass in the disk and the outer radius of the disk are two parameters that have similar effects on the interaction of the disk with the binary, i.e., increasing the mass of the disk can be counteracted by also increasing the outer radius. Therefore, in this model the outer radius of the disk is fixed at 500 AU, and only the disk mass is considered as a free parameter. The R$_{\rm out}$ = 500 AU is based on observation of the dense regions of CB disks around post-AGB systems. 

The stability of mass-loss depends both on the geometry of the system and on the reaction of the donor star. Generally it is assumed that when mass is removed from an RGB star with a convective envelope, the star will respond by increasing in radius, which is also called case-B mass-loss \citep{Thomas1977}. Stability criteria are usually linked to the mass-ratio of the system \citep[e.g.][]{Soberman1997}, and would find almost all mass-loss on the RGB to be unstable. However, many improvements and adjustment have been made, especially in the treatment of the donor star's response to mass-loss \citep[e.g.][]{Ge2010, Pavlovskii2015, Pavlovskii2017}. Currently MESA does not automatically check the stability of mass-loss. 

As can be seen, many of the processes describing stable-RLOF are heavily parameterised. Moreover, most of these parameters are currently completely unconstrained. To improve the predictive power of binary evolution models, it is necessary to find constrains for the input parameters, or provide a theoretical framework replacing them.

\section{Model improvements and observables}\label{sec:improvements}

\subsection{Mass-loss fractions}
The mass-loss fractions ($\alpha, \beta$ and $\delta$) affect how mass is lost from the system, and which fraction of it will end up in the CB disk. The main difference between these fractions is the amount of angular momentum that the mass-loss removes from the system. For mass lost to the CB disk ($\delta$), part of this angular momentum can be returned to the system through the Lindblad resonances, but that is a very small amount. SPH simulations can be used to link the mass-loss fractions to the geometrical properties of the system, such as component masses, orbital period and eccentricity. For example, the Oil on Water SPH code has been used to determine what happens during mass-loss phases in white dwarf - neutron star binaries \citep{Church2012, Bobrick2017}. Similar studies could be performed in the case of RGB+MS binaries to find a description of the mass-loss fractions in terms of system properties. This can be validated by comparing the results of binary population synthesis (BPS) codes implementing this description with the observed sdB sample in terms of orbital period distribution.

Furthermore, the amount of accreted matter can be estimated by studying the rotational velocity of the companions. By assuming that the systems were circularised and synchronised before the onset of mass-loss, and attributing the increase of rotational velocity of the secondary to mass accretion during RLOF, the amount of accreted mass can be calculated \citep{Vos2017b}. Furthermore, the evolved state of some companions argues in favour of low mass accretion rates \citep[e.g.][]{Vos2012}.

\subsection{CB-disk parameters}
The main observable effect of a CB disk is increasing the eccentricity of the orbits. There is a small effect on the orbital period, but this is negligible compared to the effect of the mass-loss fractions. The effect of a CB disk on binary evolution could be improved by including a self consistent treatment of the CB disk \citep[e.g.][]{Rafikov2016}. BPS models could then be compared to the observed eccentricity distribution to validate the model. 
  
There are currently no direct observations of CB disks around sdB binaries. Their existence however is supported by the existence of CB disks around AGB stars \citep[e.g.][]{Hillen2017} and by the discovery of dusty post RGB binaries \citep{Kamath2016}. Direct observations of dust around sdB binaries, with for example ALMA, could be used to further constrain the properties of CB disks.

\subsection{Stability of RLOF}
The stability criteria for RLOF needs to be able to explain the observed population of wide sdB binaries. While it is not generally possible to determine the mass ratio of the wide sdB binaries at the start of the interaction phase, some conclusions can be made. In the case of a companion that has evolved off the MS, the initial mass ratio of the system has to have been very close to 1. If this was not the case, the companion would not have had enough time to evolve. On the other hand, systems with dwarf companions will have an initial mass ratio much less then unity. By observing at which period ranges these two types of systems exist, estimations on stability criteria can be made \citep{Vos2017a}.

Furthermore, the current mass-ratio of the wide sdB binaries in combination with the spectral type of the companions can also be used to test the stability criteria.

\section{Conclusions}\label{sec:conclusions}
A significant fraction of the low mass binaries will undergo mass-loss on the RGB, either entering in a common envelope phase or by undergoing stable RLOF. Both interactions are still not understood from first principles. With our project to study wide sdB+MS binaries we aim at improving the current understanding of RLOF on the RGB. In this article we have provided an overview of the currently used model for RLOF and discussed several methods to improve it. Furthermore, we have shown how updates of the different interaction physics in the model could be tested by comparing the updated models to observable properties of wide sdB binaries.

\bibliographystyle{aa}
\bibliography{bibliography}

\end{document}